\documentclass[final,3p,times]{elsarticle}
\usepackage{hyperref}
\usepackage{amsmath}
\usepackage{amsfonts}
\usepackage{amssymb}
\usepackage{bm}
\usepackage[numbers]{natbib}
\usepackage{tabulary}
\usepackage{placeins}

\journal{Current Opinion in Structural Biology}

\begin{document}

\begin{frontmatter}



\title{Towards deep learning sequence-structure co-generation for protein design}


\author[inst1,int]{Chentong Wang}
\affiliation[inst1]{organization={School of Life Sciences, Westlake University},%Department and Organization
            city={Hangzhou},
            state={Zhejiang},
            postcode={310024}, 
            country={China}}
\author[inst3]{Sarah Alamdari}
\author[inst3]{Carles Domingo-Enrich}
\author[inst3]{Ava P. Amini}
\author[inst3,cor1]{Kevin K. Yang}

\ead{yang.kevin@microsoft.com}
\fntext[int]{Work done during an internship at Microsoft Research.}

\fntext[cor1]{Corresponding author.}
\affiliation[inst3]{organization={Microsoft Research},%Department and Organization
            city={Cambridge},
            state={MA},
            postcode={02142}, 
            country={USA}}

\begin{abstract}
Deep generative models that learn from the distribution of natural protein sequences and structures may enable the design of new proteins with valuable functions. While the majority of today's models focus on generating either sequences or structures, emerging co-generation methods promise more accurate and controllable protein design, ideally achieved by modeling both modalities simultaneously. Here we review recent advances in deep generative models for protein design, with a particular focus on sequence-structure co-generation methods. We describe the key methodological and evaluation principles underlying these methods, highlight recent advances from the literature, and discuss opportunities for continued development of sequence-structure co-generation approaches.
\end{abstract}


% \begin{highlights}
% \end{highlights}


\end{frontmatter}

% \linenumbers

\section{Introduction}
\label{sec:intro}
 
\begin{figure}[ht]
    \centering
    \includegraphics[width=\textwidth]{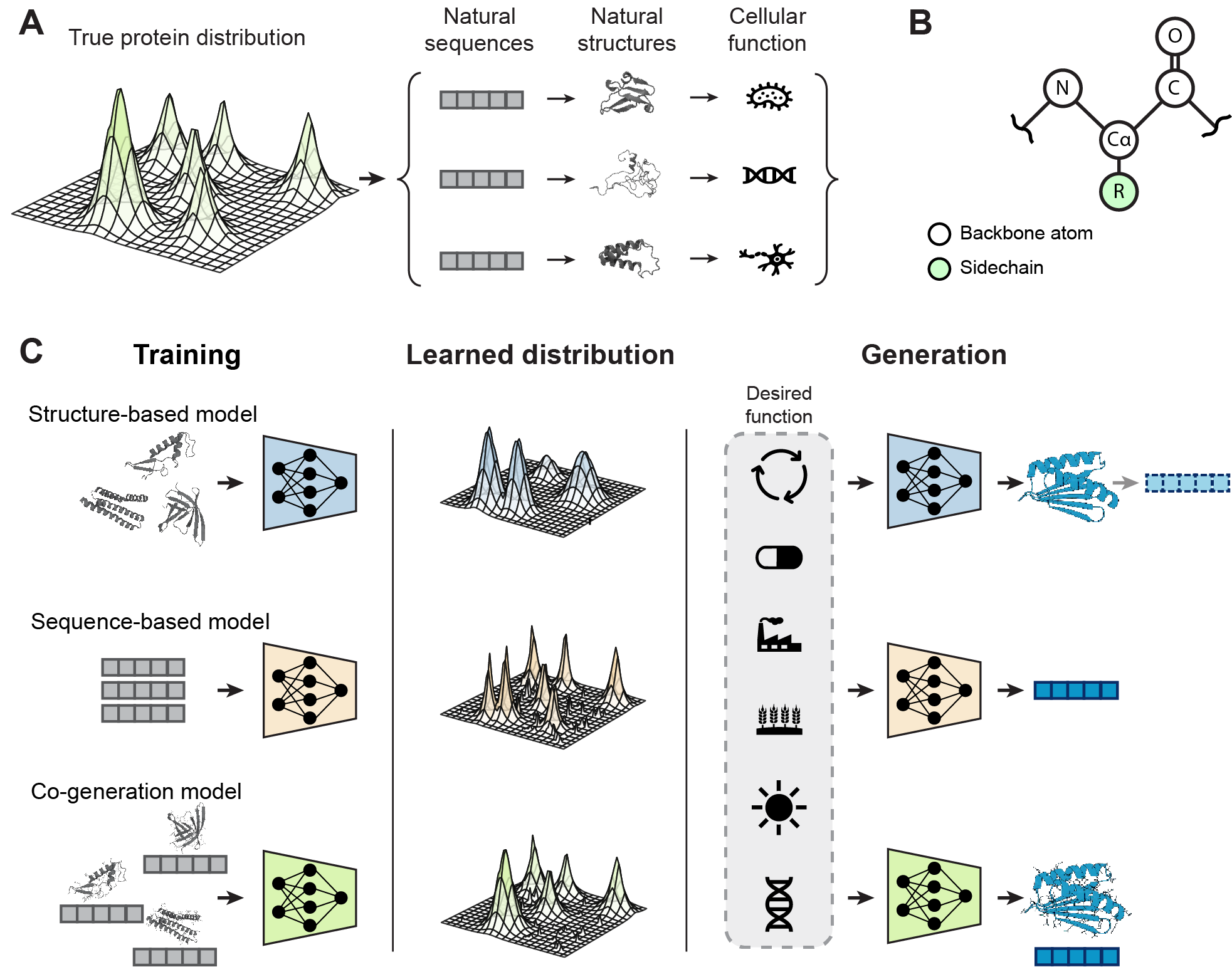}
    \caption{\textbf{Generative modeling for protein design. }\textbf{A} Natural proteins are (biased) samples from the distribution of possible sequences. Each protein's sequence determines its structure, and sequence and structure together mediate function. \textbf{B} The backbone atoms are common to all amino acid residues, while the identity of each residue is determined by the identity of the atoms in its side chain (R). \textbf{C} Overview of structure-based models, sequence-based models, and co-generation models for protein design. }
    \label{fig:overview}
\end{figure}

Deep generative models trained on natural biomolecular diversity have the potential to generate valid and novel proteins to help solve modern-day challenges, for example as effective therapeutics or engineered enzymes~\cite{strokach2022deep,wu2021protein,lovelock2022road}. 
% The function of a protein is directly determined by its three-dimensional (3D) structure, which in turn is dictated by its amino acid sequence. \textit{De novo} protein design aims to create entirely new proteins from scratch rather than modifying existing ones. 
Proteins are biomolecules comprised of a \textbf{sequence}\footnote{Bolded terms are defined in \ref{app:gloss}.} of amino acid \textbf{residues}, and this sequence dictates how a protein assembles into a 3-dimensional structure. Sequence and structure together mediate both natural and artificial protein function (Figure~\ref{fig:overview}A). Impressively, only 20 natural amino acids are needed to describe the chemical and structural diversity of all natural proteins. Each of these amino acids share a set of common \textbf{backbone} atoms, differing only by the composition of their \textbf{side chain} atoms (Figure~\ref{fig:overview}B)~\cite{branden2012introduction}. As a result, a protein's sequence can be easily described by a single string of single-letter amino acid codes. When available, structural characterization of proteins also provides 3D coordinates of the backbone and side chain atoms that comprise their structures. 

Because of the complexity of the protein sequence-structure landscape, computational methods have been developed to enable the engineering or \textit{de novo} design of new proteins. Traditional physics-based protein design methods have successfully created a variety of functional proteins~\cite{cao2022design,berger2024preclinical}; however, physics-based methods are constrained by the inaccuracies of empirical energy functions~\cite{Alford2017Rosetta} and by their dependence on existing protein structures or predefined topologies~\cite{huang2016coming}. 
In contrast, deep learning approaches that learn from large-scale protein sequence or structure databases enable the exploration of a vast and largely untapped design space, leading to the creation of proteins with novel sequences, folds, and desired functions~\cite{wu2021protein,ovchinnikov2021structure,notin2024machine,kortemme2024novo}. This capability has opened up exciting possibilities in various applications such as designing novel enzymes with tailored catalytic activities~\cite{yeh2023novo,lauko2024computational}, engineering high-affinity protein binders for therapeutic purposes~\cite{watson2023novo,bennett2023improving}, and even creating entirely new protein topologies with the potential for unprecedented functionality~\cite{verkuil2022language,Ingraham2023chroma}.

Deep generative modeling is a powerful approach in which a neural network is trained on samples from the distribution of natural protein sequences or structures in order to propose new ones. 
While the ultimate goal is to generate proteins conditioned on a specific function, practitioners often first train unconditional models and then repurpose or fine-tune models for conditional generation.
Most current deep learning methods for protein design aim to generate functional proteins either by first generating a structural backbone and then designing a sequence that will fold into that structure (structure-based design), or by directly generating a protein sequence (sequence-based design). These methods have complementary strengths and weaknesses. Structure-based methods leverage more semantically rich information and are most effective when the desired function is mediated by a small structural \textbf{motif}, but structural data is limited and biased, leading to less diverse learned distributions. Meanwhile, sequence-based methods generalize more easily to functions that rely on larger or multiple domains, are mediated by a structural ensemble, or involve intrinsically-disordered regions. However, while sequence data is more plentiful and not biased towards regions with well-defined structures, in sequence-based design the limited understanding of structural constraints and lower-quality data may lead to noisier learned distributions. 

To overcome the limitations of generating sequence or structure alone, researchers are developing sequence-structure co-generation methods, which aim to model and generate both modalities simultaneously. Reasoning about sequence and structure throughout generation and fully leveraging the available data should result in a more accurate learned distribution and more plausible generations.
This review aims to provide a comprehensive overview of recent advances in deep generative models for protein design, with a particular focus on sequence-structure co-generation methods. We begin by introducing recent advances in structure-based and sequence-based generation, laying the foundation for understanding the unique advantages and challenges of co-generation (Figure~\ref{fig:overview}C). 
Throughout the review, we describe key principles, highlight useful examples from the literature, and discuss areas that would benefit from future methodological improvements.

% Recent advancements in structure prediction~\cite{jumper2021highly} and the development of diffusion models in image generation~\cite{ho2020denoising} have significantly accelerated progress in structure-based protein design. Meanwhile, generative models in discrete space\cite{austin2023structured} and the scaling laws observed in large language models\cite{kaplan2020scaling} have led to promising results in sequence-based protein design. These distinct approaches encompasses the ability of both unconditional generation and conditional design for functional proteins.

% Given the successes in both modalities, in this review, we will first introduce the recent advances in structure-based and sequence-based generative models. Finally, we will discuss the innovations and challenges associated with the synergy of co-generating structure and sequence, a promising approach that still faces significant technical hurdles. Several recent reviews have covered technical discussions of detailed generative algorithm, and we will not repeat them here.

\section{Single-modality deep generative models of proteins}
\subsection{Structure-based methods}
% Recent advancements in structure prediction~\cite{jumper2021highly} and the development of diffusion models in image generation~\cite{ho2020denoising} have significantly accelerated progress in structure-based protein design. 

Structure-based protein design typically involves two stages: first generating a protein backbone structure, and then finding a compatible amino acid sequence. This process can be represented as:
\begin{equation}
P(\text{seq},\text{structure}) = P(\text{structure}) \times P(\text{seq}|\text{structure}).
\label{eq:structure_design}
\end{equation}
% By prioritizing scaffold formation in the initial step, this approach ensures the resulting amino acid sequence meets the desired structural and functional goals. 
Generative models of structure learn $P(\text{structure})$ and then rely on a separate \textbf{backbone-conditioned sequence design} model~\cite{dauparas2022mpnn,hsu2022learning,yang2023masked} to compute $P(\text{seq}|\text{structure})$.
% Once the two-stage design is complete, structure prediction models\cite{jumper2021highly,baek2021rosettafold} are used to validate whether the designed sequence is likely to fold into the generated three-dimensional structure. 
There are three notable approaches to modeling $P(\text{structure})$: hallucination\footnote{While hallucination methods generate both sequence and structure, they utilize only the generated structure and do not directly model $P(\text{structure})$, instead using structure prediction and backbone-conditioned sequence design models to match sequences to generated structures.}, denoising diffusion probabilistic models (DDPMs), and flow matching.

Hallucination methods~\cite{Anishchenko2021,Christoffer2021trrosetta,wang2022scaffolding,Frank2023sequence} begin with a random sequence and use discrete optimization or backpropagation through a structure prediction model~\cite{jumper2021highly,baek2021rosettafold} to find sequences that are predicted to fold to a stable structure with high confidence. Additional loss terms can be used to specify other structural constraints. Hallucinated sequences are often adversarial, so a high-quality sequence for the generated structure is re-designed using a backbone-conditioned sequence design method.
Denoising diffusion probabilistic models (DDPMs; \cite{ho2020denoising,song2021denoising,song2021scorebased}) are generative models trained to iteratively refine noised data into realistic protein backbones~\cite{Anand2022ProteinSA,trippe2023ProtDiff,Wu2024foldingdiff,Lee2023proteinsgm,Ingraham2023chroma,watson2023novo,yim2023framediff,wang2024proteus,lin2024genie2}. 
Flow matching models~\cite{lipman2023flow,liu2023flow,albergo2023building,albergo2023stochastic,chen2024flow,huguet2024sequence} directly learn a mapping from a simple probability distribution to the target distribution of interest, without simulating the stochastic corruption and generation process.
We provide a brief overview of diffusion and flow matching models in~\ref{app:ddpm}.

The goal for structure-based methods is to produce designable backbones that are distinct from each other and from natural structures.
% Structure-based methods are evaluated \textit{in silico} by assessing the \textbf{designability}, \textbf{diversity}, and \textbf{novelty} of the generated backbones. 
\textbf{Designability} refers to the ability of current methods to find a sequence that is predicted to fold to the generated structure, and is estimated by predicting sequences from the generated backbone, passing those sequences to a structure prediction model, and measuring how self-consistent the generated and predicted structures are. 
Hallucination methods produce highly-designable and diverse backbones without additional training; however, sampling from diffusion and flow-matching models is much faster. 
Flow-matching models claim superior generation quality over diffusion models while being more computationally efficient; however, there is a lack of rigorous ablation studies demonstrating the superiority of flow matching over diffusion models in protein structure generation.
% The diffusion-based model RFDiffusion~\cite{watson2023novo} leverages a pretrained structure prediction module to achieve very strong \textit{in silico} designability scores supported by high wet-lab success rates for unconditional generation. 
Structure-based models can be used to generate functional proteins via \textbf{motif scaffolding}, in which the structural information for a functional motif is used to conditionally generate a structural scaffold to hold the motif in place~\cite{wang2022scaffolding,watson2023novo}, or via conditioning on text descriptions of function~\cite{Ingraham2023chroma,dai2024toward}. 
Structure-based models have demonstrated success across a range of tasks, including unconditional and topology-constrained monomer design, enzyme design, and binder design~\cite{watson2023novo}.

% In addition 

% and high wet-lab success rates across a range of tasks, including unconditional and topology-constrained monomer design, symmetric oligomer generation, enzyme active site scaffolding, the design of metal-binding proteins and protein binders. Its capabilities have been further extended in two recent studies: one demonstrated de novo single-domain antibody design by fine-tuning RFdiffusion and Rosettafold2 on antibody data~\cite{Bennett2024antibody}, while the other enabled de novo binder design for intrinsically disordered regions by incorporating target structure prediction alongside binder generation within RFdiffusion~\cite{Liu2024idrbinder}.
However, they are limited by the relatively small number of experimental structures (approximately 225,000 structures in the PDB~\cite{berman2000protein}) available for training. 
Additionally, these models struggle to access functions mediated by disordered regions, which are under-represented in the PDB, as they rely on dynamics rather than a single stable structure~\cite{alamdari2023protein}. Moreover, current structure generation methods fall short in modeling fine-grained structures that include side chain conformations, which are critical for catalysis and protein-protein interactions, as the identities and placement of the side-chain atoms are tied to both the amino acid types and backbone structure. 
% However, the reliance on limited structural data and the challenges in modeling structur dynamics and side-chain conformations highlight the need for further innovation in this field.

\subsection{Sequence-based methods}

Sequence-based methods seek to generate plausible proteins directly by training on the distribution of natural sequences. Protein sequence models are especially promising for functions that are not mediated by a single stable structure, those that do not fall explicitly into a binding-based paradigm, and those that involve intrinsically disordered regions.
To date, the most notable approaches for sequence-based protein design are autoregressive and diffusion models. 
Autoregressive models predict each amino acid residue from the previous ones~\cite{madani2023large,hesslow2022rita,ferruz2022protgpt2,nijkamp2023progen2,truong2023poet,sgarbossa2024protmamba}, while discrete diffusion methods are trained to iteratively convert corrupted sequences into realistic protein sequences~\cite{alamdari2023protein,wang2024diffusion,hayes2024simulating}. 
In absorbing state, or masking, diffusion~\cite{hoogeboom2022autoregressive}, residues are replaced by a special mask token, and the model is trained to predict the identities of the true residues given the unmasked residues~\cite{alamdari2023protein,wang2024diffusion,hayes2024simulating}. 
Protein masked language models~\cite{hayes2024simulating,yang2024convolutions}, which learn to reconstruct a fixed proportion of masked residues from the unmasked residues, can be considered a special case of absorbing state diffusion when used generatively. 
Discrete diffusion denoising probabilistic models are trained to iteratively denoise a sequence that has been randomly mutated~\cite{austin2023structured,alamdari2023protein}.
Continuous diffusion models can also be applied to generate protein sequences by operating in the latent space of another pretrained model~\cite{meshchaninov2024diffusion,zhang2023pro}. 

% Proteins generated from sequence-based models are notoriously difficult to evaluate \textit{in silico}. evaluation

For unconditional generation, autoregressive models have been shown to outperform masking and discrete diffusion models at learning the sequence distribution~\cite{alamdari2023protein,serrano2024protein,cheng2024training}.
However, masking diffusion models have demonstrated the ability to perform motif scaffolding directly in sequence space~\cite{alamdari2023protein,wang2024diffusion}. 
Sequence-based models have also been used to generate sequences with a specified function by conditioning on the desired structure~\cite{verkuil2022language}, fine-tuning on a family of natural proteins with the desired function~\cite{subramanian2023unexplored,madani2023large,ruffolo2024design}, conditioning on the desired reaction class~\cite{munsamy2024conditional,nicolini2024fine}, and using gradients from a function prediction model~\cite{zhang2023pro,gruver2024protein}. 
% Unlocking Guidance for Discrete State-Space Diffusion and Flow Models
% Simplified and Generalized Masked Diffusion for Discrete Data

Compared to structure-based methods, sequence-based methods have much more available training data -- UniProt contains about 300 million sequences~\cite{uniprot2019uniprot}. 
However, sequence is further removed from function. 
% and the sequences available are biased by evolution and by a preference for easy-to-study organisms, which in turn biases the resulting generative models~\cite{ding2024protein}. 
As a result, even when they exhibit low sequence similarity to individual natural proteins, functional sequences generated from sequence-based generative models are often chimeras of several natural proteins instead of being truly \textit{de novo}~\cite{ruffolo2024design,hayes2024simulating}, although it is possible to sample stable examples of novel structures from sequence-based models~\cite{verkuil2022language}.
While they are aware of residue identities and therefore implicitly learn side chain distributions and relationships, current sequence generation methods do not reason directly about the placement of side chain atoms.

\section{Sequence-structure co-generation}

\begin{figure}[ht]
    \centering
    \includegraphics[width=0.7\textwidth]{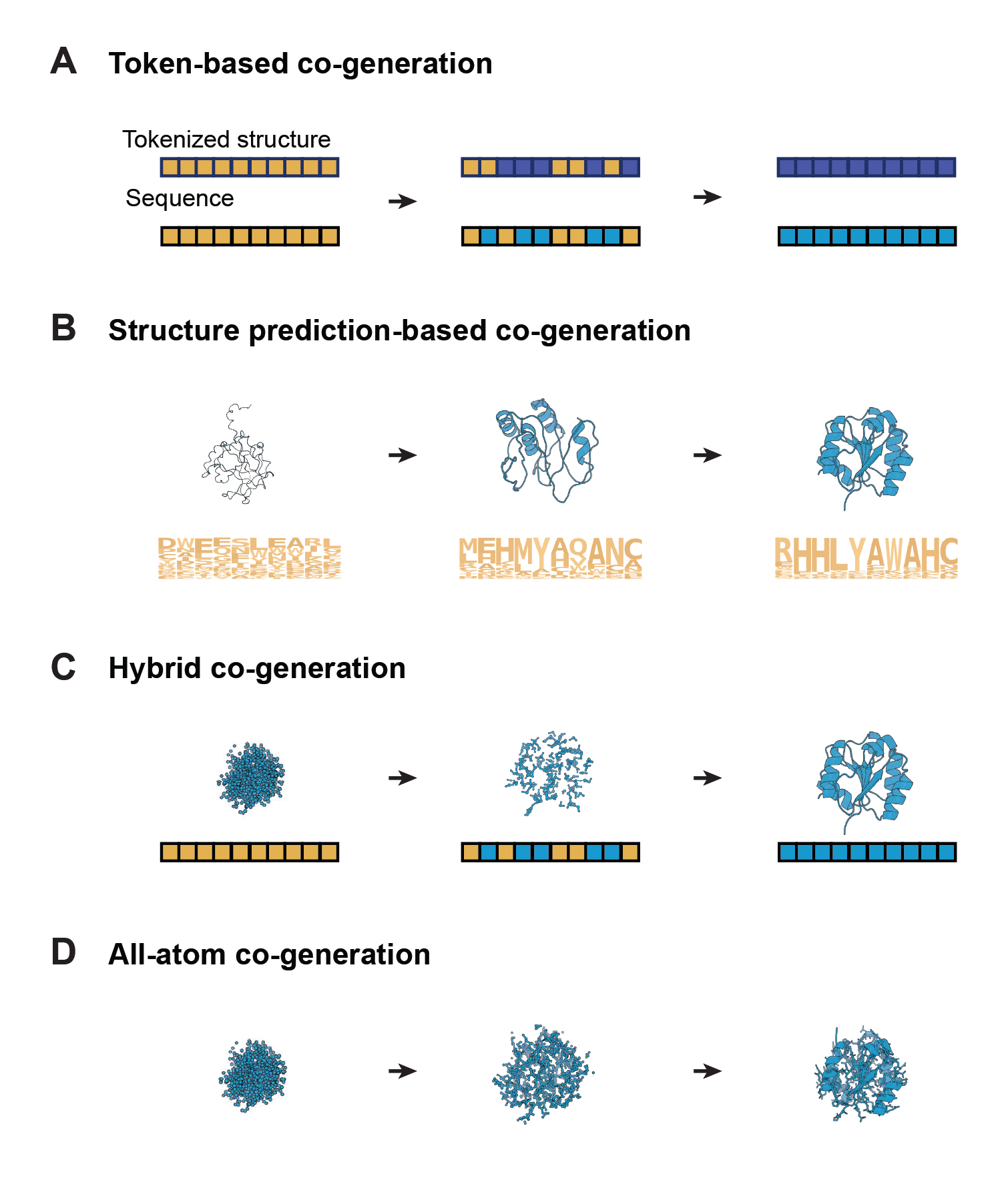}
    \caption{\textbf{Overview of current sequence-structure co-generation methods.} \textbf{A} Token-based co-generation models convert structure to discrete tokens. \textbf{B} Structure prediction-based co-generation models make sequence continuous and guide generation with a structure prediction model. \textbf{C} Hybrid co-generation models use the same neural network to generate a discrete sequence and a continuous backbone structure. \textbf{D} All-atom co-generation models directly generate the identity and location of every atom in the structure.}
    \label{fig:cogen}
\end{figure}
% cogeneration with dna: https://arxiv.org/abs/2401.06151

% Transition that addresses the strengths and weaknesses of str/seq based methods. Talk about difficulty of data augmentation using predicted structures here. Talk about lack of physical knowledge from sequence only / PLM scaling / limits coevolution
% While structure-based protein design methods excel at incorporating precise structural motifs to guide functional design, they are constrained by the limited availability and diversity of experimentally determined structures. In contrast, sequence-based models, trained on vast amounts of sequence data, offer broader exploration of sequence space but lack the ability to steer generation directly towards desired functions, often relying on fine-tuning with curated sequences. Limiting the scope of generation to known protein families and functions~\cite{ruffolo2024proselm}.
Sequence-structure co-generation aims to combine the strengths of structure- and sequence-based modeling and to unlock atomistic control over protein design. 
% a little bit of history about non deep-learning methods
Rosetta's ``flexible backbone design" protocol was one of the earliest to adopt this philosophy~\cite{saunders2005rosetta}, by starting with a coarse-grained designed backbone and then alternating between designing a sequence given the current backbone and optimizing the backbone given the current sequence. While this approach has been successful in applications such as designing small protein binders~\cite{cao2022design}, it relies on hand-crafted energy functions for both backbone and sequence design instead of learning the underlying distributions from data; in fact, using neural network models for structure prediction or backbone-conditioned sequence design in this pipeline greatly improved success rates~\cite{bennett2023improving}.
An early attempt at co-generation using deep learning focused on ``inpainting'' tasks, in which a region of a protein is masked out and regenerated jointly with both structure and sequence~\cite{wang2022scaffolding}.
% Summary of deep learning methods
However, the true potential of co-generation lies in generating entire proteins. Recent advances have enabled models to generate both sequence and structure from scratch. The central challenge of co-generation is how to pass information between the sequence and the structure. Current strategies for mixing sequence and structure information within the model include tokenizing the structure, leveraging a structure prediction model, mixing discrete and continuous processes, and all-atom generation (Figure~\ref{fig:cogen}).

% is it necessary to make a seperate subsection for latent space generation? maybe we can make it in disccusion as there are few paper 
%\subsection{Latent diffusion}
%CHEAP amy lu

\subsection{Token-based co-generation models}
% Co-design models vary significantly in how they represent protein structures and sequences, impacting their architecture and generative process. 
One strategy for co-generation is to first convert the continuous structure to a sequence of discrete structure tokens using techniques such as vector-quantized variational autoencoders (VQ-VAEs)~\cite{vqvae2017} (Figure~\ref{fig:cogen}A). Concatenating the sequence and structure tokens then enables language-model style architectures to be repurposed for co-generation.  
Foldseek~\cite{vanKempen2024foldseek} applied VQ-VAEs to tokenize the structural environment around each residue in order to use efficient sequence-based homology search methods for structural homology search and clustering. 
This tokenization scheme was later applied to masked language models to train joint representations of sequence and structure~\cite{heinzinger2023bilingual,Su2023saprot}. 
In ESM3~\cite{hayes2024simulating}, the authors train their own structure tokenizer and show that their multimodal architecture enables the unconditional generation of high-quality proteins, motif scaffolding, and, by incorporating function tokens, function conditioning. 

Co-generation methods that tokenize structure can take advantage of the extensive neural network machinery for modeling sequences of tokens and can train on datasets where not every sequence is paired with a structure.
However, they do not directly model side-chain atom positions during generation, and instead use the VQ-VAE decoder to convert structure tokens to all-atom coordinates after generation.
Furthermore, while previous work has shown that autoregressive language models outperform masked language models for generation, there is not yet an autoregressive joint sequence-structure token model. 
Finally, the quality of these models is upper-bounded by the fidelity and expressiveness of the structure tokenization, and the design choices here remain mostly unexplored. 

\subsection{Structure prediction-based co-generation models}
Instead of discretizing structural information, ProteinGenerator~\cite{lisanza2023joint} employs a diffusion process in continuous sequence space to generate protein sequence and structure jointly. By leveraging the pretrained RoseTTaFold2 network~\cite{baek2021rosettafold} for structure prediction, the model predicts and conditions on the corresponding structure at each diffusion step (Figure~\ref{fig:cogen}B). This approach circumvents the need to explicitly model the structure distribution.  ProteinGenerator has shown experimentally-validated ability to scaffold flexible peptides and to design proteins capable of adopting multiple conformations, a challenging task that requires simultaneous optimization on distinct conformational states. Unlike previous methods that need predefined desired conformations before sequence optimization\cite{Praetorius2023hinge}, ProteinGenerator achieves this by tying the sequences from several generation processes together with distinct secondary structure guidance. While theoretically applicable to a broader range of design tasks, ProteinGenerator lags behind RFdiffusion in motif scaffolding and unconditional generation of larger proteins, suggesting limitations associated with the absence of explicit training on structure generation. However, generation in sequence space may enable the use of sequence-based function predictions for conditioning.

\subsection{Hybrid co-generation models}

Hybrid co-generation models combine the discrete sequence-generation process with the continuous backbone-generation process without altering the fundamental properties of the underlying data (Figure~\ref{fig:cogen}C).
CarbonNovo interleaves, at every time step, a diffusion model for backbone structure generation with a structure-conditioned Markov random field for sequence generation~\cite{ren2024carbonnovo}. 
In contrast, DiffAb~\cite{luo2022antigenspecific} and MultiFlow~\cite{campbell2024generative} add a discrete sequence diffusion or flow-matching process to a diffusion- or flow-matching-based backbone generation model. 
These methods use the same neural network and a shared representation to learn two intertwined generative processes. 
% add a possibly corrupted residue to the input for each position, and predict the true amino acid at every position. 

These hybrid models simply and efficiently combine advances in discrete and continuous generative models without dealing with the complexity and possible loss in fidelity of tokenizing structure or embedding sequence. 
However, they require paired sequence-structure data and do not directly model side-chain atom positions during generation.
Furthermore, it is unclear whether adding sequence information to backbone generation methods is sufficient to generate more plausible backbone structures, as these methods do not demonstrate substantial gains in designability and diversity over RFdiffusion. 
Besides motif scaffolding, function conditioning has yet to be explored with these methods, although the inclusion of sequence information should enable the use of sequence-based function predictors for conditioning.

\subsection{All-atom co-generation models}
Direct generation of all-atom protein structures (i.e., the complete set of protein backbone and side chain atoms) implicitly enables co-generation, as the side-chain atoms directly represent the amino acid identities that make up a protein's sequence (Figure~\ref{fig:cogen}D). 
%Given the intrinsic link between side-chain atom identity and amino acid type, generating all-atom structures necessitates coupled sequence generation. 
The central challenge in all-atom co-generation lies in the model's requirement to flexibly handle a varying number and type of side-chain atoms at each step. At the initial sampling stage, the backbone, sequence, and number of atoms are all unknown. 
Protpardelle~\cite{Chu2024Protpardelle} employs Euclidean diffusion for both backbone and side-chain atoms. To circumvent the variable-residue challenge, Protpardelle tracks all 73 possible non-hydrogen atoms for each residue.
%This set encompasses all heavy atoms across all amino acid types. 
At each diffusion step, the predicted backbone guides sequence design, followed by the denoising of corresponding side-chain atoms in the \texttt{atom73} coordinates. 
To bypass the need for decoding the amino acid identity in an intermediate step, a more recent work, P(all-atom)~\cite{qu2024p}, proposes \texttt{atom14} coordinates as a unified atomic representation, as the largest amino acid has 14 non-hydrogen atoms. The model then applies simple Euclidean diffusion on these \texttt{atom14} coordinates without intermediate sequence decoding. Because each amino acid type has a distinct collection of side chain conformations, the model learns to implicitly assign an amino acid identity to each residue, based on specific conformational patterns within the \texttt{atom14} coordinates. The sequence is thus decoded after generation, and the generated \texttt{atom14} coordinates are mapped back to the all-atom atomic index for the assigned amino acid.

All-atom generation models hold great promise for protein design, as they promise precise control over side-chain specific interactions throughout the generative process. This level of control is critical for tackling challenging tasks such as enzyme and antibody design. All-atom generation naturally extends to the co-generation of non-peptide interacting molecules, such as DNA, RNA, or small molecules. However, the high dimensionality of all-atom representations poses significant computational hurdles, necessitating the development of more efficient model architectures~\cite{Abramson2024}. In addition, it is unclear how to leverage sequences without structures in an all-atom model. Overall, this remains a largely unexplored but fertile area for future research.

\section{Discussion and outlook}

This review examines the early efforts at using deep learning to co-generate protein sequence and structure. While these efforts have yielded promising results, there is still room to improve the consistency of the generations, to fully leverage unpaired sequences for training and evaluation, and to test the capabilities of co-generation for real-world applications.

\textbf{Self-consistency} between the generated sequence and structure remains challenging. 
Current techniques model sequence and structure jointly but ultimately sample them independently given the model outputs. Furthermore, some methods use an absorbing state diffusion process for the sequence without iterative refinement -- once unmasked, an amino acid identity cannot be corrected as the generated structure is refined. 
Therefore, current co-generation models produce less designable structures than two-stage structure-based methods. 
For example, co-generated sequences from Protpardelle are less self-consistent with a generated backbone structure than sequences designed from the same structure using ProteinMPNN~\cite{Chu2024Protpardelle}. 
By training on the most self-consistent sequences generated by ProteinMPNN from native and generated backbones, Multiflow's~\cite{campbell2024generative} co-generation method achieves comparable self-consistency to an equivalent two-stage process. 
While ESM3~\cite{hayes2024simulating} theoretically allows simultaneous generation of sequence and structure, its GFP design pipeline alternates between generating structure tokens based on functional motifs and then optimizing the sequence based on the structure, repeated for multiple cycles.

While generative models that unify both modalities would ideally combine the strengths of structure- and sequence-based methods, current approaches often overemphasize structure in their architecture, data, and evaluations, potentially limiting their ability to explore the full design space where sequence plays a more dominant role in determining function -- crucial for tasks like conformation switching and enzyme design.
Many co-generation methods are either dependent on a pretrained structure prediction module or are direct adaptations of architectures designed for backbone generation. 
All the co-generation methods we cover require paired sequence-structure data. Methods that use structure when available but default to sequence-only when not remain underexplored, and thus the scope of available sequence data has not been fully exploited for co-generation. 
While using \textit{in silico}-predicted structures seems like a possible way to bridge this gap, naive augmentation has widely been observed to harm generation quality\cite{Su2023saprot,huguet2024sequence}. 

This bias towards structure is perhaps most apparent in our reliance on structural self-consistency for evaluating generation quality and in the dominance of structure-focused tasks, such as binder design or motif scaffolding, for downstream applications. These evaluations favor sequences that fold into stable structures (or, conversely, very stable backbones), limiting exploration of applications where function-driven dynamics are more important than rigidity. Self-consistency metrics are inherently constrained by the capabilities of current structure prediction and backbone-conditioned sequence design models. 
The structure prediction models commonly used in these pipelines can be insensitive to the effects of sequence variation that are not seen in nature. For example, for generated enzymes, AlphaFold prediction confidence does not correlate with activity or expression~\cite{johnson2024computational}.  
In addition, ProteinMPNN achieves high native sequence recovery for fewer than half of the high-confidence predicted structures in AlphaFoldDB~\cite{lin2024genie2}.
These observations suggest that predicting sequence from backbone is easier for some proteins, while predicting structure from sequence is easier for others. 
This property motivates the development of performant co-generation models that can learn from both sequence and structure simultaneously. 
By generating both in tandem, methods for protein sequence-structure co-generation could overcome the complexities of mapping between these individual modalities and ultimately enable a broader, more diverse range of downstream applications for generative protein design.

\FloatBarrier
% For example, Genie2 found that fewer than half of the high-confidence predicted structures in AFDB can be inverse folded by ProteinMPNN, This finding underscores that relying solely on AlphaFold2's confidence estimates is insufficient for selecting high-quality structures suitable for training structure generation models.
% In the realm of inverse folding, expanding the training data beyond the PDB dataset is equally desirable to enhance the capability of sequence design models to handle diverse and novel protein structures. However, directly training models on AlphaFold-predicted structures can lead to overfitting, making them overly sensitive to backbone noise and often producing sequences that are not experimentally viable[ESMIF]. Therefore, while leveraging the vastness of AFDB is appealing, it's crucial to address the inherent discrepancies between predicted and experimentally determined structures. 

\section*{Acknowledgements}

The authors thank Sergey Ovchinnikov for discussions on hallucinations and Alex Chu for discussions on Protpardelle. 

%\appendix

%\section{Sample Appendix Section}
%Lorem ipsum dolor sit amet, consectetur adipiscing elit, sed do eiusmod tempor section \ref{sec:sample1} incididunt ut labore et dolore magna aliqua. Ut enim ad minim veniam, quis nostrud exercitation ullamco laboris nisi ut aliquip ex ea commodo consequat. Duis aute irure dolor in reprehenderit in voluptate velit esse cillum dolore eu fugiat nulla pariatur. Excepteur sint occaecat cupidatat non proident, sunt in culpa qui officia deserunt mollit anim id est laborum.

\bibliographystyle{elsarticle-num-annotate} 
\bibliography{refs}





% \end{thebibliography}
\newpage
\appendix

\section{Glossary}
\label{app:gloss}
\begin{table}[h]
\centering
\begin{tabular}{|m{3.5cm}|p{10cm}|}
\hline
\textbf{Term} & \textbf{Description} \\ \hline
All-atom co-generation & A co-generation strategy that directly generates the complete protein structure, including all backbone and side chain atoms. \\ \hline
Backbone & A protein's backbone consists of its N, C${\alpha}$, C, and O atoms. All amino acids contain these atoms, so the backbone does not depend on the identities of the amino acids. \\ \hline
\parbox[c]{3.5cm}{Backbone-conditioned\\sequence design} & The design of an amino acid sequence that is likely to fold into a predefined three-dimensional protein backbone. \\ \hline
Designability & The degree to which a backbone structure can be realized by an amino acid sequence. \\ \hline
Diversity & The degree to which generated structures or sequences differ from each other. \\ \hline
Hybrid co-generation & A co-generation strategy where the protein structure is represented in its native continuous form and the protein sequence is represented in its native discrete form. \\ \hline
Motif & A specific region in a protein with a conserved sequence or structural pattern, often associated with a specific function. \\ \hline
Motif scaffolding & The design of a new protein that incorporates a specific motif as a core element, with the surrounding protein context designed to support and enhance the motif's function. \\ \hline
Novelty & The degree to which generated structures or sequences differ from naturally occurring proteins. \\ \hline
Residue & An individual amino acid within a protein. \\ \hline
Self-consistency & The agreement between a generated sequence and its corresponding generated structure. A self-consistent sequence is likely to fold into the intended structure, typically assessed using structure prediction tools. \\ \hline
Sequence & The linear order of amino acid residues in a polypeptide chain, typically listed from the N-terminal to the C-terminal. \\ \hline
Side chain & The variable chemical group (R) attached to the alpha carbon of an amino acid, determining its identity and chemical properties. \\ \hline
\parbox[c]{2.5cm}{Structure prediction-based co-generation} & A co-generation strategy that leverages a structure prediction network to jointly generate the protein sequence and structure, where the structure is predicted from the sequence. \\ \hline
\parbox[c]{2.5cm}{Token-based\\co-generation} & A co-generation strategy where the protein structure is converted into discrete tokens, enabling the use of language-model-like architectures. \\ \hline
\end{tabular}
\caption{Glossary of terms used in the review.}
\label{tab:glossary}
\end{table}

\newpage
\section{Diffusion models and flow matching} \label{app:ddpm}
Flow matching and diffusion models are the state of the art methods for generating data in continuous spaces, such as Euclidean spaces and manifolds \cite{chen2024flow}. Their goal is to generate realistic samples that resemble the training dataset. These techniques are used in fields ranging from text-to-image, text-to-audio, and text-to-video models, to the modeling of protein backbones. In simple terms, these methods create new data by transforming random noise into structured data that matches the distribution of the data of interest.

Training flow matching models involves two main objects: the reference flow and the generative flow. The reference flow is constructed by drawing random variables from an initial distribution, $\bar{X}_0 \sim p_0$, which represents random noise, and from the distribution of the data that we want to model, $\bar{X}_1 \sim p_1 = p_{\mathrm{data}}$. In practice, $\bar{X}_1$ is sampled from the training dataset. Then, the reference flow is defined as an interpolation path $\bar{X}_t$ between the noise sample $\bar{X}_0$ and the data sample $\bar{X}_1$. For example, in Euclidean space it is standard to choose $p_0$ to be the multivariate standard Gaussian, and a linear interpolation path $\bar{X}_t = \beta_t \bar{X}_0 + \alpha_t \bar{X}_1$ for $t$ between 0 and 1, where the functions $\alpha_t$ and $\beta_t$ are chosen appropriately.
% The process begins with random variables drawn from an initial distribution, $\bar{X}_0 \sim p_0 = \mathcal{N}(0, I)$, which represents random noise. We also have $\bar{X}_1$, which follows the distribution of the data we want to model. The evolution of these random variables over time is described by the reference flow $\bar{X}_t = \beta_t \bar{X}_0 + \alpha_t \bar{X}_1$,
% Here, the functions $\alpha_t$ and $\beta_t$ ensure that at time $t = 0$, the system starts from noise ($\alpha_0 = \beta_1 = 0$), and at time $t = 1$, the variables align with the data distribution ($\alpha_1 = \beta_0 = 1$). 

% Diffusion models and flow matching algorithms construct 
The generative flow $X_t$ is a process 
% $\bm{X} = (X_t)_{t \in [0,1]}$ 
that begins from the same noise as the reference flow ($X_0 \sim %\mathcal{N}(0, I)
p_0$) and evolves to produce new samples that look like the real data. 
The key idea is that, at any time $t$, the distributions of the generative flow $X_t$ and the reference flow $\bar{X}_t$ are identical, so by the final time $t = 1$, $X_1$ follows the same distribution as the data.
In flow matching, the generative process follows an ordinary differential equation (ODE) defined as:
\begin{equation} \label{eq:FM_ode} 
\frac{\mathrm{d}X_t}{\mathrm{d}t} = v_{\theta}(X_t,t), \qquad X_0 \sim p_0, \end{equation}
Here, $v_{\theta}(\cdot, \cdot)$ is a velocity field, which is usually modeled using neural networks, such as a U-Net \citep{ronneberger2015unet} or diffusion transformer architecture \citep{peebles2022scalable}. The goal is to optimize the network so that it matches the behavior of the reference flow over time, which is achieved by minimizing the difference between the predicted velocity and the actual derivative of the reference flow \citep{lipman2023flow,albergo2023building}. Once $v_{\theta}$ has been trained, generating an artificial sample involves drawing a random variable from $p_0$ and solving the ODE \eqref{eq:FM_ode} using the numerical method of choice, such as the Euler scheme. 

Diffusion models \citep{ho2020denoising,song2021denoising,song2021scorebased} are a specific form of flow matching on Euclidean spaces, where the reference flow uses a multivariate standard Gaussian as the noise distribution, and linear interpolation paths with coefficients $\alpha_t = \sqrt{\bar{\alpha}_t}$ and $\beta_t = \sqrt{1-\bar{\alpha}_t}$. In this case, instead of directly learning the velocity field, the neural network is a noise predictor $\epsilon_{\theta}$ which predicts the original noise $X_0$. The training objective is to minimize the error in this noise prediction. Diffusion models appeared before flow matching, but they are being superseded by the latter due to improvements in sample generation quality and efficiency.
% The generative process for diffusion models is described by a stochastic differential equation (SDE):

% \begin{equation} \label{eq
% } \mathrm{d}X_t = \bigg( \frac{\dot{\bar{\alpha}}{t}}{2\bar{\alpha}{t}} X_t - \bigg( \frac{\dot{\bar{\alpha}}{t}}{2\bar{\alpha}{t}} + \frac{\sigma(t)^2}{2} \bigg) \frac{\epsilon_{\theta}(X_{t},t)}{\sqrt{1-\bar{\alpha}{t}}} \bigg) , \mathrm{d}t + \sigma(t) \mathrm{d}B_t, \qquad X{0} \sim \mathcal{N}(0,I), \end{equation}

% In this equation, $\dot{\bar{\alpha}}_{t}$ is the derivative of $\bar{\alpha}_t$, and $\sigma(t)$ is a noise schedule, which typically controls the randomness in the generative process.

\end{document}